\documentclass[
reprint,amsmath,amssymb,aip]{revtex4-1}

\usepackage{graphics}
\usepackage{amsmath}
\usepackage{amsfonts}
\usepackage{latexsym}
\usepackage{amsbsy}
\usepackage{epstopdf}

\usepackage{color}

\usepackage{ulem} 

\newcommand{\dyadic}[1]{{#1}
\setbox0=\hbox{$\scriptstyle\leftrightarrow$}
   \setbox2=\hbox{$#1$}
   \dimen0=.5\wd0 \advance\dimen0 by-.5\wd2
   \advance\dimen0 by-\wd0
   \kern\dimen0
{^{\hbox{$\scriptstyle\leftrightarrow$}}}}

\begin{document}

\title{Millimeter Wave Detection via Autler-Townes Splitting in Rubidium Rydberg Atoms}
\thanks{This work was partially supported by DARPA's QuASAR program. Publication of the U.S. government, not subject to U.S. copyright.}
\author{Joshua A. Gordon}
\email{josh.gordon@nist.gov}
\author{Christopher L. Holloway}
\affiliation{National Institute of Standards and Technology (NIST), Electromagnetics Division,
U.S. Department of Commerce, Boulder Laboratories,
Boulder,~CO~80305}
\author{Andrew Schwarzkopf}
\author{Dave A. Anderson}
\author{Stephanie Miller}
\author{Nithiwadee Thaicharoen}
\author{Georg Raithel}
\affiliation{Department of Physics, University of Michigan, Ann Arbor, MI 48109}

\date{\today}

\begin{abstract}
In this paper we demonstrate the detection of millimeter waves via Autler-Townes splitting in $^{85}$Rb Rydberg atoms. This method may provide an independent, atom-based, SI-traceable method for measuring mm-wave electric fields, which addresses a gap in current calibration techniques in the mm-wave regime. The electric-field amplitude within a rubidium vapor cell in the WR-10 wave guide band is measured for frequencies of \linebreak 93 GHz, and 104 GHz.  Relevant aspects of Autler–-Townes splitting originating from a four-level electromagnetically induced transparency scheme are discussed. We measure the E-field generated by an open-ended waveguide using this technique. Experimental results are compared to a full-wave finite element simulation.

\vspace{7mm}
\end{abstract}

\maketitle

The detection of millimeter waves (mm-waves) has proven useful for a broad range of applications, including weapons stand-off detection \cite{security}, aeronautics \cite{aero}, remote sensing \cite{remote} , next generation wireless communications \cite{comm} and, stand-off human vital sign monitoring \cite{vital} to mention a few. Each of these applications may be appropriately suited to one or more of the myriad of sensor types available. 


Traceability for both electric field and power measurements at these frequencies is through power. Typical sensors are Schottky diodes, bolometers, and calorimeters, all of which are traceable to calorimeter measurements. Calorimeter and bolometer measurements are traceable to DC voltage and resistance measurements \cite{microwavepwr}. The sensors and the calorimeters measure power at a reference plane (often a connector) of a rectangular wave guide. If a direct measurement of the electric field is desired, then models and or measurements, such as near-field antenna pattern techniques \cite{nf}, are needed to obtain the relationship between the desired electric field and the power at the reference plane. However, specifying reliable models and performing antenna pattern measurements becomes difficult at mm-wave frequencies because the mechanical tolerances and repeatability of such components may vary such that they are a significant fraction of the operating wavelength. For these reasons we are investigating more direct and traceable techniques for  mm-wave electric field and power calibrations. Here we present on an atomic-based technique which allows direct measurement of the magnitude of the electric field, $|E|$, at mm-wave frequencies via Autler-Townes (AT) splitting in Rydberg atoms. This splitting is inversely proportional to Planck's constant, $\hbar$ providing a link to the SI. We obtain data for a range of electric field levels in the WR-10 band (75-110 GHz) at 93.71 GHz and 104.77 GHz. Fundamentally this technique also lends itself to measurements beyond 110 GHz, which may address a current lack of traceable methods for calibrating mm-wave systems above 110 GHz. Furthermore, it does not rely on \textit{a priori} knowledge of an antenna pattern for determining the electric field. In addition this technique has many novel properties useful for measurements of $|E|$ that we have recently reported on, such as extremely large bandwidth\cite{overview} (1-500 GHz),\linebreak sub wavelength imaging \cite{sublamb} and two-photon AT interactions at microwave frequencies for potential use in high power microwave sensing \cite{2phot}

\begin{figure}
\centering
\scalebox{.45}{\includegraphics*{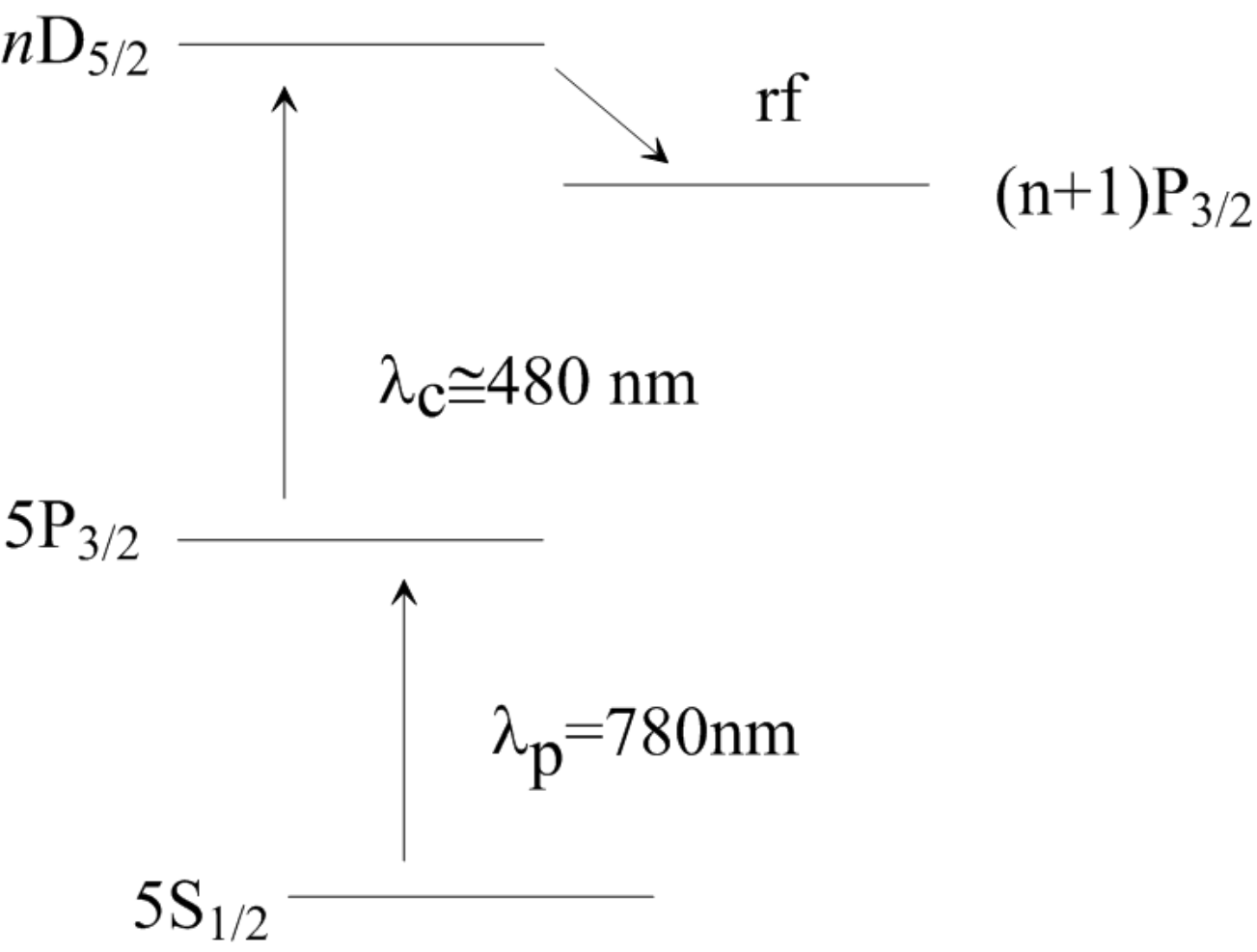}}
\caption{Energy levels used in the experiment for generating Autler-Townes splitting from $^{85}$Rb Rydberg atoms.}
\label{4level}
\end{figure}

Rydberg atoms have a single valence electron in a highly excited state, where the principal quantum number is typically $\textit{n}>10$. The dipole moment, $\wp$, in Rydberg atoms scales as $\textit{n}^2$, and at the $n$ required for a mm-wave transition ($n\sim30$), it can be several orders of magnitude greater than for a ground state atom $\sim1000ea_0$, where $\textit{e}$ is the electron charge and $a_0$ is the Bohr radius. Therefore Rydberg atoms can have significant response to mm-wave electric fields. 
In this paper we will focus on the $^{85}$Rb isotope of rubidium excited to $n=28,29$, between the $nD_{5/2}$ to $(n+1)P_{3/2}$ manifolds, corresponding to Rydberg transitions in the WR-10 mm-wave band. As has been well established in the literature \cite{gallagher} the energy, $W(n^*)$, of Rydberg states may be modeled as,

\begin{equation}
W(n^*)=-\frac{R_{Rb}}{(n^*)^2} \quad,
\label{rydberg}
\end{equation}

\noindent where $R_{Rb}$ is the Rydberg constant for the reduced electron mass in rubidium, $n^*=n-\delta$ is the effective principle quantum number determined using the quantum defect\cite{gallagher,qdefect}, $\delta$ .  The quantum defects from \cite{qdefect} were used to determine the specific mm-wave frequencies for each transition. \\
\indent Although a thorough discussion of electromagnetic induced transparency (EIT) is beyond the scope of this paper, a brief description of the phenomena is give. In a gas of rubidium atoms an incident probe laser beam experiences large absorption when tuned to the $D_2$ transition, at $\lambda_p=780.241$\,nm.  However, in the presence of a second coupling laser at, $\lambda_c\approx480$\,nm this gas will be rendered partially transparent to the probe laser and a transmission peak will result in the spectrum of the probe laser. This quantum interference effect between the ground state and states excited by the probe and coupling laser is known as Electromagnetic Induced Transparency\cite{harris}. This has been widely studied for both Rydberg atoms as well as alkali atoms at lower $n$ and was first demonstrated by Boiler et. al. . Later the effects of adding a fourth level to the EIT scheme (see Figure \ref{4level}) were theoretically investigated \cite{sharma}.  In this scheme the the transition to this fourth level is taken to be a radio frequency (RF) transition of either a hyper fine transition or a Rydberg state transition\cite{sharma}. In the case we present here, the transition to the fourth level is a mm-wave Rydberg transition. To be clear, in keeping with popular nomenclature in the literature, the use of the term RF will be used interchangeably to mean mm-wave frequencies in the rest of this paper. When Rydberg atoms are used in a four level system, the strength of the RF transition at modest electric field strengths is sufficient to transition this four level system from the EIT regime into the AT regimes \cite{at},\cite{eitcomp} where the RF electric field causes the EIT peak to split into two peaks. 

With the coupling laser on resonance, and the probe frequency swept, the EIT peak is observed to split into two equal peaks separated by the Rabi frequency, $\Omega_{RF}$, of the RF transition,

\begin{equation}
{\Omega}_{RF}=\dfrac{\wp_{RF}|E_{RF}|}{\hbar}
\label{splitting}
\end{equation}

In actuality the frequency splitting, ${\Delta}f_{probe}$, that is measured on the probe laser EIT spectrum, must be scaled by the ratio of laser wavelengths. This is in order to take into account the effects of Doppler mismatch, which occurs due to the different wavelengths of the counter propagating probe and coupling laser beams interacting with the thermal vapor (room temperature) of atoms \cite{dopplerave}. The measured splitting on the probe laser is thus related to ${\Omega}_{RF}$ in terms of the wavelengths of the coupling laser, $\lambda_c$, and probe laser, $\lambda_p$, by

\begin{equation}
{\Delta}f_{probe}=\dfrac{\lambda_c}{\lambda_p}\dfrac{\Omega_{RF}}{2\pi}
\label{deltaprobe}
\end{equation}

From (\ref{splitting}) we see that this splitting is linearly proportional to the RF electric field strength, $|E_{RF}|$, the dipole matrix element, $\wp_{RF}$ of the Rydberg RF transition, and $\hbar$. This direct relationship of the measured Rabi frequency to the electric field, the dipole matrix element and Planck${'}$s constant is at the heart of the traceability of this technique.
Sedlacek et. al. \cite{shaffer} used this technique for the $53D_{5/2}-54P_{3/2}$ Rydberg transition in $^{87}$Rb to measure the electric field strength at 14.23 GHz \linebreak  inside a vapor cell. In this paper we extend this technique for measuring electric fields in the mm-wave regime. \linebreak Frequencies in the WR-10 band of $f_0=93.71$ GHz, and $f_0=104.77$ GHz corresponding to the, $29D_{5/2}-30P_{3/2}$, and $28D_{5/2}-29P_{3/2}$ transitions respectively, are measured over a range of electric field strengths. Data are presented comparing the electric field determined via this AT splitting technique to numerical simulations performed using a three dimensional finite element approach.

\begin{figure}
\centering
\scalebox{.5}{\includegraphics*{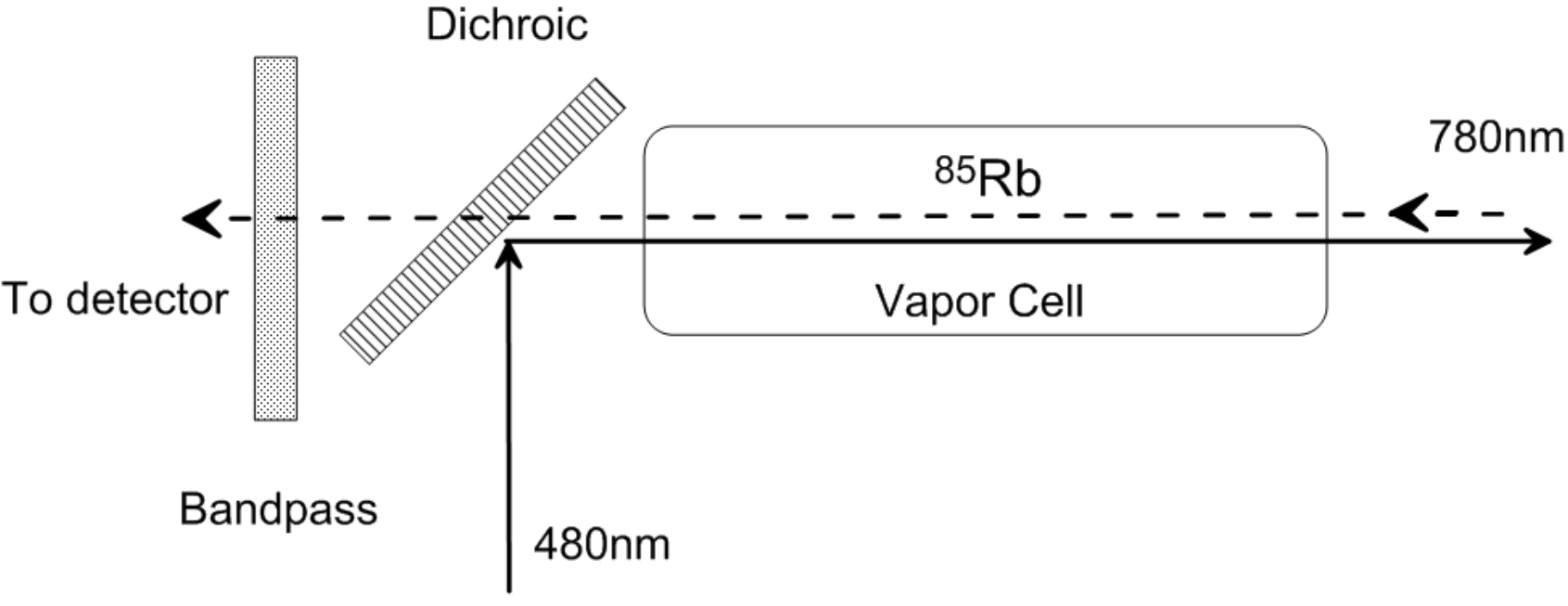}}
\caption{Experimental setup for generating EIT and AT spectra. Counter-propagating 780 nm probe laser (dotted line) and 480 nm coupling beam (solid line) are shown, as well as the Rb vapor cell, dichroic beam splitter and bandpass filter.}
\label{opticalsetup}
\end{figure}

Our experimental setup is shown in Figure \ref{opticalsetup}. Two counter-propagating lasers beams were used, the probe laser tuned to the $D2$ transition of $^{85}$Rb at \linebreak 780.241 nm  and the coupling laser tuned to approximately 480 nm for exciting Rydberg states are incident on the room temperature vapor cell. This is depicted in Figure \ref{opticalsetup}. The full-width half-max beam diameters at the center of the vapor cell for the probe and the coupling laser are 80 ${\mu}m$ and 100 ${\mu}m$ respectively. The beam powers were nominally 28 mW for the coupling laser and 100 nW for the probe laser. The line widths for both probe and coupling lasers were $\approx 1$ MHz.The probe laser does not need to be broadband tunable because it is always probing the same transition (i.e. $^{85}$Rb $D2$ line). For the coupling laser it is advantageous to have broadband tunability over a range of at least several hundred GHz to be able to optically excite a selection of Rydberg levels.

\setlength{\belowcaptionskip}{-5pt}

The mm-waves were produced using an RF signal generator (SigGen) with 0.1 Hz resolution to drive a \linebreak WR-10 6x frequency multiplier. The output of the frequency converter was coupled to a WR-10 open ended rectangular wave guide (OEG) placed approximately 140 mm from the vapor cell. The cell is a 25 mm x 75 mm hollow glass cylinder containing $^{85}$Rb vapor commonly used in saturation absorption spectroscopy. The vapor cell was mounted on a low permittivity foam block to isolate mm-wave scattering from surrounding metal optics mounts and microwave absorber was used to cover exposed surfaces of the optics bench. A variable in-line attenuator was used to vary the mm-wave power. The mm-wave power was verified for each dial position of the variable attenuator using a WR-10 power meter connected directly to the output of the OEG. Because the attenuator uses a mechanical vane to achieve attenuation, the equal dial settings did not necessarily correspond to equal steps in mm-wave power. Therefore, the power output from the OEG, $P_{OEG}$ , was calibrated using the WR-10 power meter for each frequency, and at each increment on the variable attenuator. The reflection coefficient, $|S_{11}|$,  between the OEG aperture and free space was measured on a vector network analyzer. Since the power meter is impedance matched to the OEG, the power actually leaving the OEG when coupled to free space is determined by modifying the power meter reading by $(1-|S_{11}|^2)$ so as to take in account the aperture reflection missing in the matched power meter reading.

\begin{figure}
\centering
\scalebox{.4}{\includegraphics*{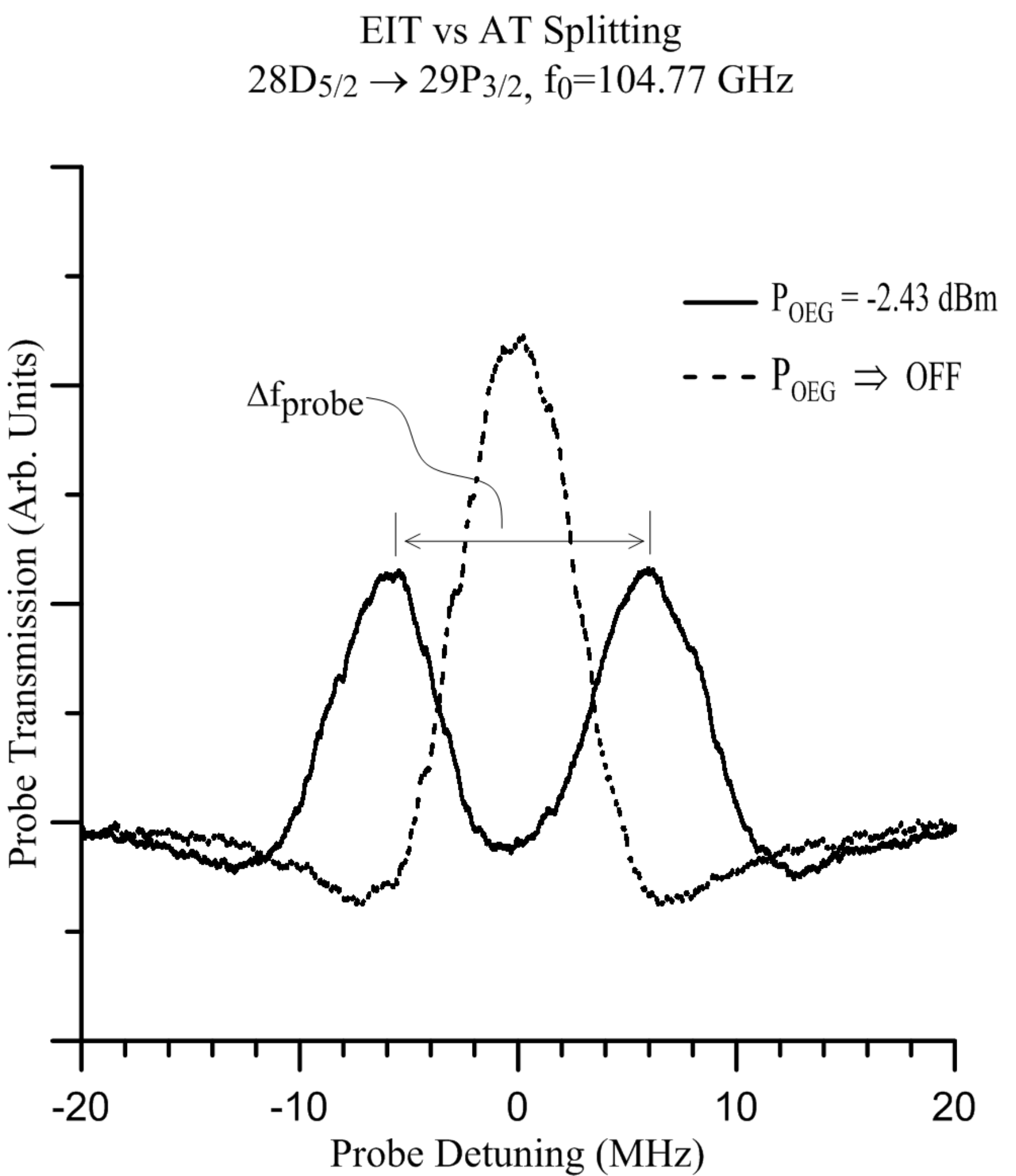}}
\caption{EIT peak with mm-wave power off and and AT splitting for $P_{OEG}=-2.43$ dBm for the $28D_{5/2}-29P_{3/2}$ transition, $f_{0}=104.77$ GHz. }
\label{atsplit}
\end{figure}

The full range of the variable attenuation was used, however the mm-wave power range at each frequency was not the same because the power produced by the mixer decreased as the frequency increases. Therefore the power range at 93.71 GHz is larger than at 104.77 GHz.\linebreak This results in a variation of achievable dynamic range between the mm-wave frequencies. The maximum power measured at the OEG aperture was -0.83 dBm at 104.77 GHz, and +1.95 dBm at 93.71 GHz.  The minimum power measured at the OEG aperture which gave unambiguous AT splitting was -11.58 dBm and -12.71 dBm, at 104.77 GHz and 93.71 GHz respectively. For each level of $P_{OEG}$ the AT signal was measured on the probe laser using a silicon photodiode and lock-in amplifier. The probe laser was separated from the coupling laser using a dichroic beam splitter followed by a 10 nm wide line filter in front of the photo diode see Figure \ref{opticalsetup}. The lock-in signal was generated by chopping the coupling laser beam using an acousto-optic modulator to produce a 30 KHz square wave modulation. With the probe laser sweeping across the Doppler spectrum of the $D2$ transition, the coupling laser was tuned to the wavelength for the desired Rydberg state. The coupling laser wavelength was determined using the $^{85}$Rb ionization energy and $D2$ transition energy from \cite{steck}, and the calculated Rydberg state energy using (\ref{rydberg}), fine-tuning was done by observing the three-level EIT signature (with mm-wave power off). Once the EIT signal was established the coupling laser was locked to a stabilizing cavity.\\
\indent To produce the AT signal, the SigGen was tuned to the mm-wave frequency that was determined again using (\ref{rydberg}) for the desired Rydberg transition. With the variable attenuator set to half of the power range, the mm-wave frequency was fine-tuned so as to result in AT peaks of equal heights. Both the laser fields and mm-waves were (linear) $\pi$-polarized and aligned so as to minimize excitation of the 3-level EIT pathway \cite{shaffervect}. The mm-wave power was then varied using equal increment dial settings on the attenuator. The Doppler-free saturation absorption spectrum of the probe laser was obtained simultaneously with the AT spectrum. This was used to calibrate the measured AT splitting from the known frequency spacing of the Doppler-free hyperfine features present in the $D2$ saturation spectrum. Equation(\ref{splitting}) was then used to determine the $\Omega_{RF}$, where the dipole moment $\wp_{RF}$, was calculated using the methods described in \cite{gallagher} and the appropriate Clebsch-Gordan coefficients. Figure \ref{atsplit} shows scans taken for the $28D_{5/2}-29P_{3/2}$ transition for the EIT signal with mm-wave power off, along with the AT signal with the power at the OEG set to -2.43 dBm.

\begin{figure}
\centering
\scalebox{.4}{\includegraphics*{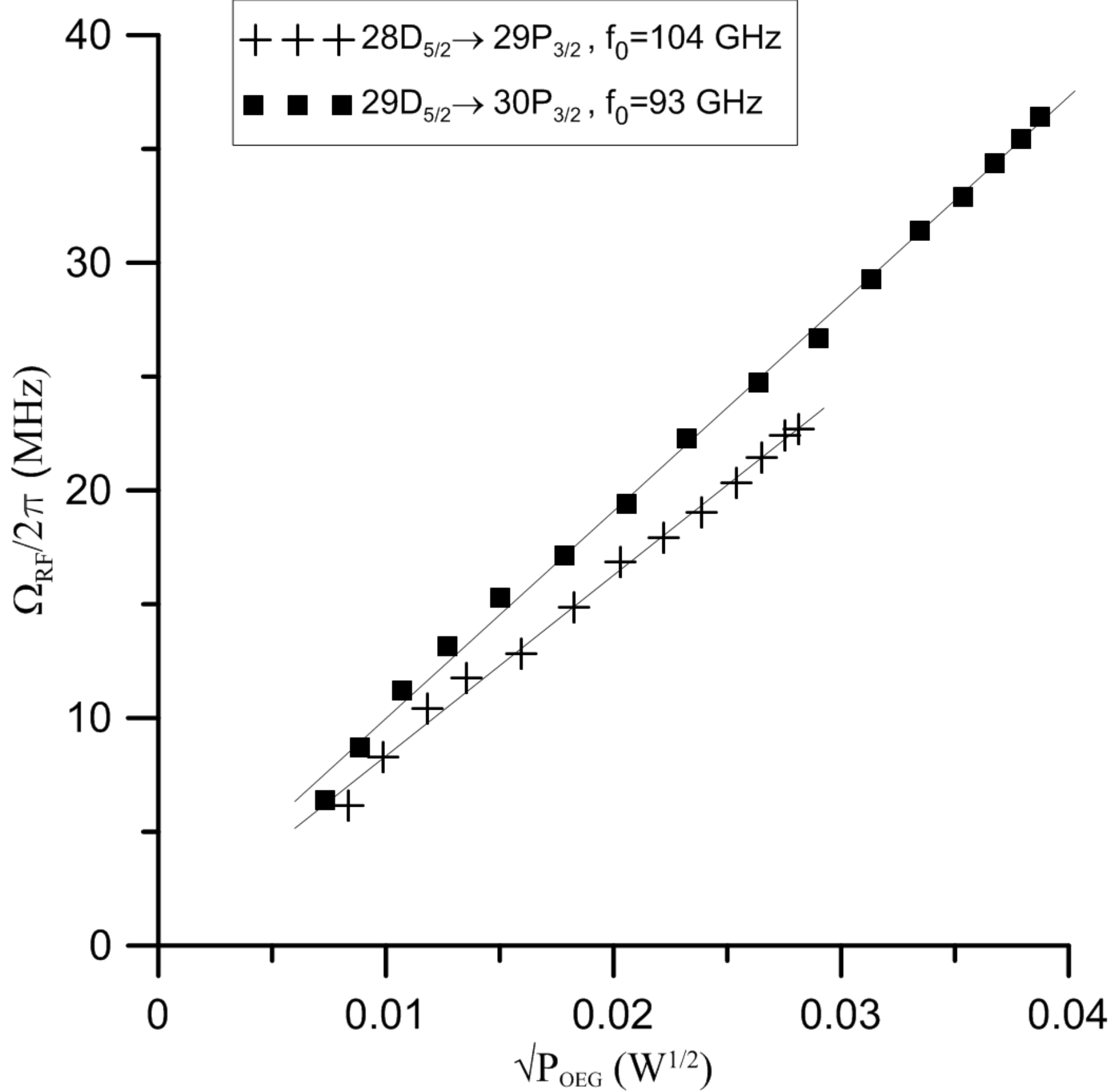}}
\caption{AT splitting in MHz versus $\sqrt{P_{OEG}}$ for the $29D_{5/2}-30P_{3/2}$, $f_0=93.71$ GHz, and $28D_{5/2}-29P_{3/2}$ $f_0=104.77$ GHz transitions. Linear fits are shown as solid lines.}
\label{splitpwr}
\end{figure}

\begin{figure}
\centering
\scalebox{.4}{\includegraphics*{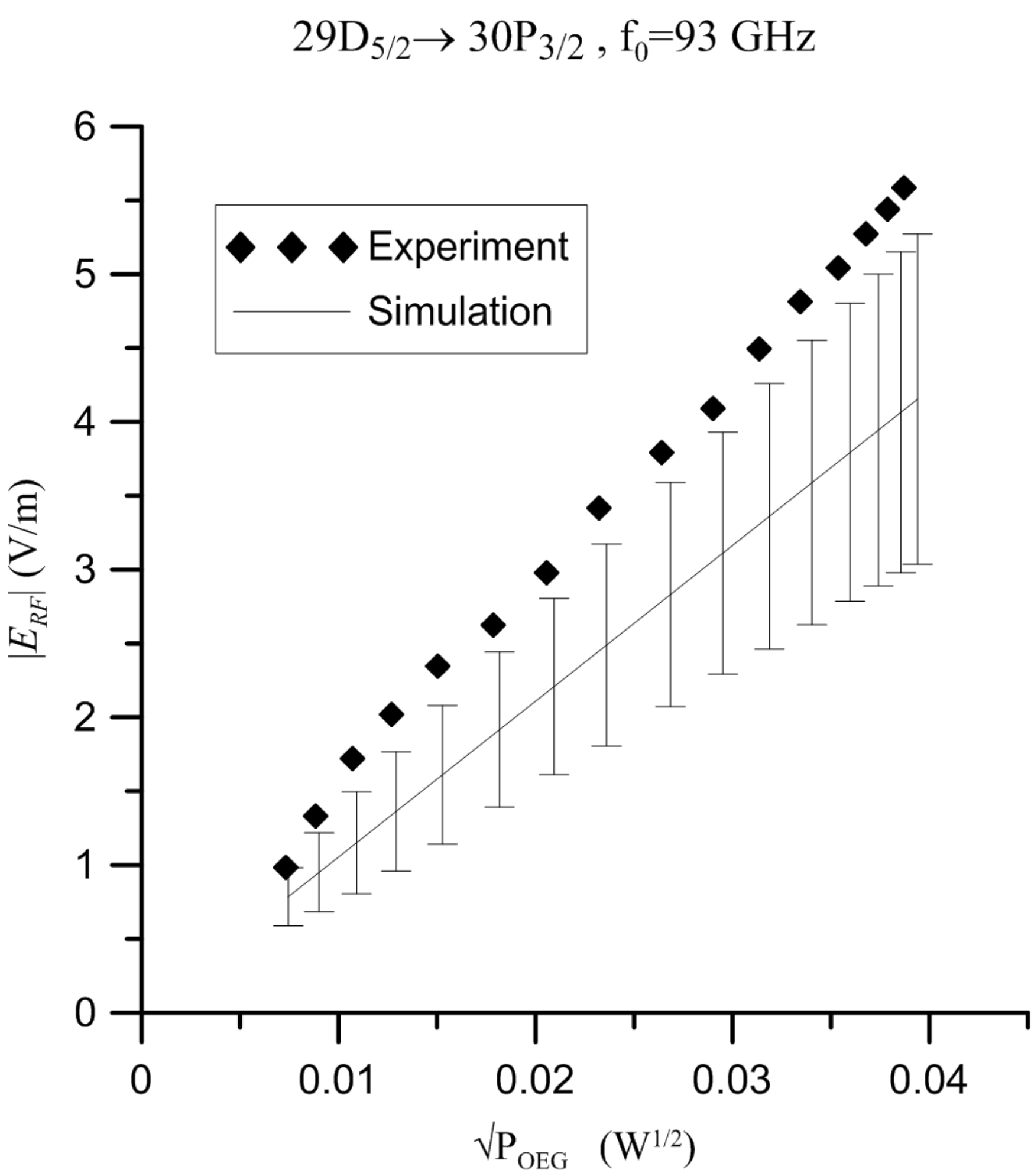}}
\caption{Electric field values at 93.71 GHz measured using the vapor cell and compared to HFSS far-field simulation. Solid line shows HFSS simulation. Error bars indicate the possible $20\%$ field variation range due to resonant mm-wave scattering effects of the vapor cell.}
\label{splitpwr93}
\end{figure}

\begin{figure}[!t]
\centering
\scalebox{.4}{\includegraphics*{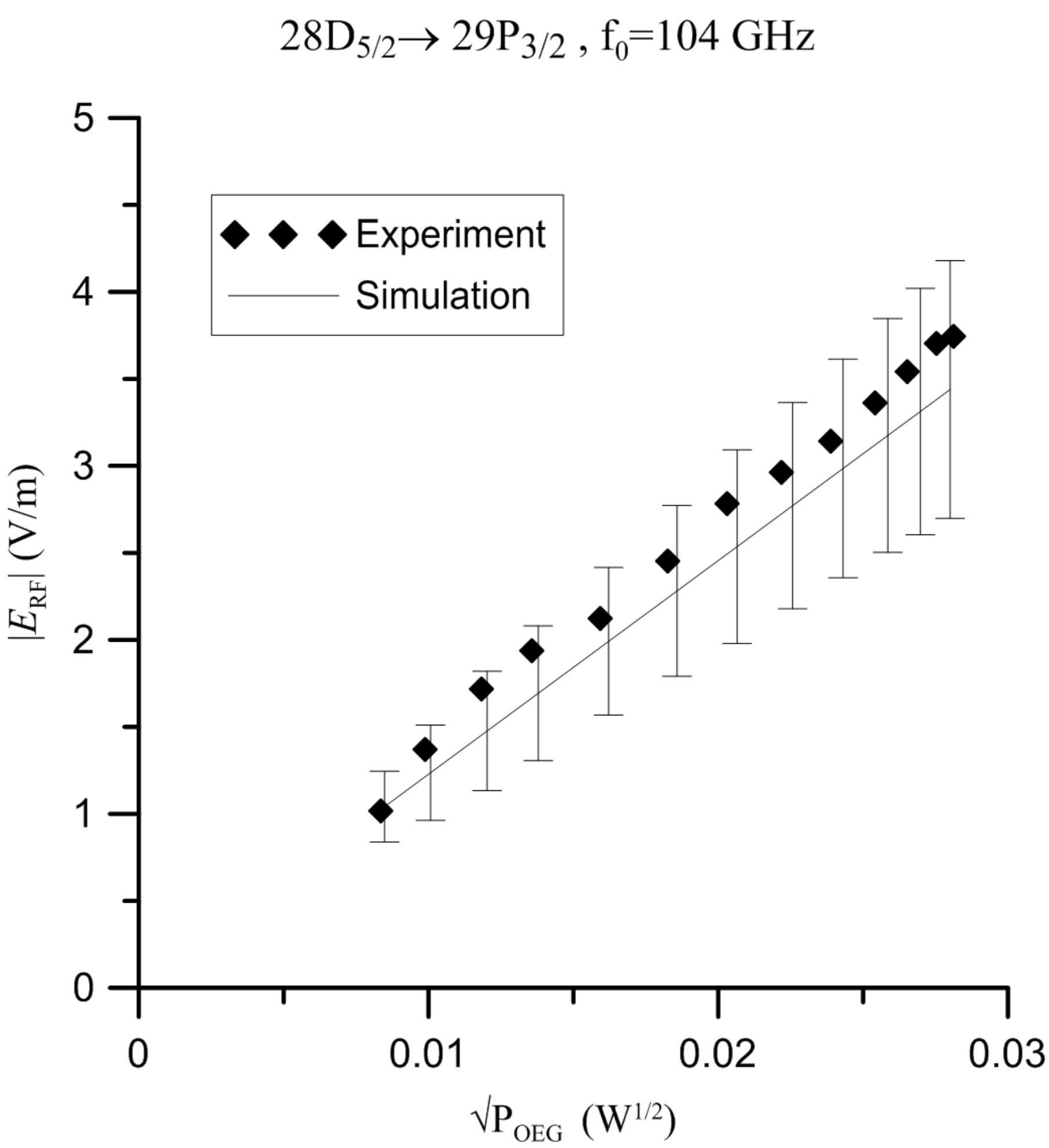}}
\caption{Electric field values at 104.77 GHz measured using the vapor cell and compared to HFSS far-field simulation. Solid line shows HFSS simulation. Error bars indicate the possible $\%20$ field variation range due to resonant mm-wave scattering effects of the vapor cell.}
\label{splitpwr104}
\end{figure}

Electric fields were using the experimental setup described above. For each power setting of the variable attenator the splitting was calculated using (\ref{splitting}) and (\ref{deltaprobe}) and compared to simulated results. At a distance of 140 mm from the aperture of the OEG, the vapor cell was well beyond the farthest far-field distance calculated for the mm-wave frequencies that were measured (i.e. $\approx$5.63 mm, the value at \linebreak 104.77 GHz). The far-field distance was calculated using the dimension of the OEG aperture diagonal and the conventional definition given in \cite{ieee}. Simulations were performed to determine the electric field radiated by the OEG using the far-field calculator in the electromagnetic finite element solver HFSS (mention of this software is not an endorsement but is only intended to clarify what was done in this work). These field values were then used to compare with those measured using the vapor cell.
\indent First, we established that the measured splitting scales linearly with the electric field as expected from (\ref{splitting}). Given that the electric field at the vapor cell is proportional to the $\sqrt{P_{OEG}}$\,, if the splitting indeed follows the behavior in (\ref{splitting}), then a linear relationship would be apparent by plotting $\Omega_{RF}$ versus $\sqrt{P_{OEG}}$. This is clearly shown in Figure \ref{splitpwr} for both frequencies. Figures \ref{splitpwr93} and \ref{splitpwr104} show a comparison of the electric field values determined from the vapor cell measurements to those obtained from the HFSS far-field simulation. The error bars in these plots show the expected range of electric field values within the vapor cell due to field variations that are present because of the  dielectric boundary of the cell. We discuss this further next. 

From Figures \ref{splitpwr93} and \ref{splitpwr104},  we see that there is a \linebreak noticeable difference between the measured and simulated electric field values. Also, the agreement of the measured electric field to simulated results is not consistent between frequencies. A strongly observable effect which alters the electric field inside the vapor cell results from standing waves set up by the dielectric boundary of the cell walls interacting with the mm-waves. The dimensions of the vapor cell used are $\approx 25$ mm x $75$ mm and the operating wavelengths are 3.22 mm and 2.88 mm at 93.71 GHz and 104.77 GHz respectively. Therefore, the vapor cell, in terms of wavelengths is $\approx7.7 \lambda$ x $23.4\lambda$ and $\approx8.7\lambda$ x $26.3\lambda$ for these two cases. Since the laser beams were not moved between measuring the two mm-wave frequencies, the observed frequency dependence is attributed to the difference in mm-wave mode structure of the vapor cell as a result of the difference in wavelength between the 93.71 GHz and 104.77 GHz frequencies. As the wavelength changes, the standing wave structure changes, and thus the electric field amplitude at the location of the laser beams in the vapor cell will depend on the mm-wave frequency. We have reported on this effect in detail in \cite{sublamb}, where we show the ability to image these standing waves at extreme sub-wavelength resolution using AT splitting at both microwave (17.04 GHz) and mm-wave (104.77 GHz) frequencies. From imaging these standing waves we determined a $\pm20\%$ variation about the mean field strength as a function of measurement location in the vapor cell at 104.77 GHz. The error bars in Figures \ref{splitpwr93} and \ref{splitpwr104} indicate this $\pm20\%$ variation for the measurements we present here. This perturbing effect of the electric field by the presence of the dielectric vapor cell is something we are currently addressing.

In this paper we demonstrate the detection of millimeter waves via Autler-Townes splitting in $^{85}$Rb \linebreak Rydberg atoms. This method may provide an independent, atomic-based, SI-traceable method for measuring mm-wave electric fields, which addresses a gap in current calibration techniques in the mm-wave regime. The electric field amplitude within a rubidium vapor cell in the WR-10 waveguide band was measured for frequencies of 93.71 GHz, and 104.77 GHz. Experimental results are presented where we measure the far-field electric field generated by an open ended waveguide using this technique. A comparison to far-field electric field values obtained from a finite element simulation is made. The experimentally observed scaling behavior follows closely the expected linear behavior of Autler-Townes splitting. The  electric fields measured agree to within $\pm20\%$ of the far-field simulations due to standing wave effects.         

\section{Acknowledgments}

Special thanks to David R. Novotny, and Galen H. Koepke of the Electromagnetics Division at NIST, Boulder for assistance with equipment.\\

\end{document}